\documentclass[a4paper]{jpconf}
\usepackage{graphicx}
\usepackage{multicol}
\usepackage{color}
\def\be{\begin{equation}}
\def\ee{\end{equation}}
\def\bac{\begin{array} {c}}
\def\ea{\end{array}}

\def\ssGL{{\scriptscriptstyle GL}}
\def\ssE{{\scriptscriptstyle E}}
\begin{document}

\hspace{8cm} FTUAM-12-122, IFT-UAM/CSIC-12-136

\title{Superconductivity, Superfluidity and Holography}

\author{Alberto Salvio}

\address{Scuola Normale Superiore and INFN, Piazza dei Cavalieri 7, 56126 Pisa, Italy

\vspace{0.2cm}
Departamento de F\'isica Te\'orica, Universidad Aut\'onoma de Madrid and \\ 
Instituto de F\'isica Te\'orica IFT-UAM/CSIC, Cantoblanco, 28049 Madrid, Spain

 }
\ead{alberto.salvio@uam.es}

\begin{abstract}
This is a concise review of holographic superconductors and superfluids. We highlight some predictions of the holographic models
and the emphasis is given to physical aspects rather than
to the technical details, although some references to understand the latter are systematically
provided. We include gapped systems in the discussion, 
motivated by the physics of high-temperature superconductivity. In order to do so we consider a compactified extra dimension (with radius $R$), 
or, alternatively,  a dilatonic field. The first setup can also be used to model cylindrical superconductors;
when these are probed by an axial magnetic field a universal property of holography emerges: while for large $R$ 
(compared to the other scales in the problem) non-local operators are suppressed, leading to the so called Little-Parks periodicity, the opposite limit 
shows non-local effects, e.g. the uplifting of the Little-Parks periodicity. This difference corresponds in the 
gravity side to a Hawking-Page phase transition.
\end{abstract}

\section{Introduction: motivations, background material and conclusions}\label{introduction}
Holographic dualities have become useful tools to explore strongly coupled theories; in particular, in the last few years 
systems at finite temperature $T$ and density of a U(1) charge have been extensively studied. The main motivation
comes from condensed matter physics (for a review see \cite{Hartnoll:2009sz}):
if the U(1) charge is the electric charge and the U(1) symmetry 
is spontaneously broken one identifies these systems with superconductors (SCs);
when the U(1) symmetry is global, instead, the spontaneously broken phase corresponds to 
superfluidity.

One of the main ideas behind the application of holography to superconductivity is the hope to shed light on the physics of unconventional
SCs (such as high-$T$ SCs), which are not completely described by the weakly coupled theory: the 
Bardeen-Cooper-Schrieffer theory. Cuprate high-$T$ SCs, the prototype of unconventional materials, are obtained
by doping a Mott insulator and as such they typically show in the ($T$ versus ``dopant concentration'') phase diagram an insulator phase close 
to the region where the U(1) symmetry is spontaneously broken \cite{doping}. This calls for a way to describe both conductor/SC and 
insulator/SC transitions. Since any insulating material eventually conducts if it is probed by a strong enough
external current, this in turn requires a method to study gapped phases. 

In this article we will briefly review holographic SCs \cite{Hartnoll:2008vx,Domenech:2010nf} and superfluids (SFs)
\cite{Herzog:2008he,Domenech:2010nf}. The emphasis will be given to  physical aspects rather than to the technical details,
for which we will refer to the original papers.
Given that all SF systems can be considered as SCs in the limit of non-dynamical electromagnetic (EM)
gauge fields, for the sake of definiteness, we will use the terminology adopted in the literature of superconductivity,
rather than that of superfluidity (unless otherwise stated). We will focus on the best understood version of holographic duality, the 
anti de Sitter/conformal field theory (AdS/CFT) correspondence. The discussion of the previous paragraph then motivates us 
to introduce  a conformal symmetry breaking in the infrared (IR), which is needed to have a gap.
Moreover, since much of the interesting physics of cuprate SCs is layered, we shall restrict our analysis to 
2+1 dimensional systems, dual to gravitational theories on four dimensional spacetimes.
Although beyond the scope of this review, it is useful to mention that other regions of cuprate phase diagrams
require a departure from the theory of Fermi liquids, a challenging property which can be realized in holography \cite{Hartnoll:2009ns}.

 In the rest of this section we shall introduce the concepts and general properties of superconductivity which are needed to
 understand the holographic results discussed in the following sections, providing at the same time the reader with
 the organization of the paper and the conclusions. We assume the minimal field content to describe superconductivity: 
 a charged scalar $\Phi_{\rm cl}$, responsible for the breaking of the  U(1) symmetry and a gauge field $a_\mu$. 
 The quantum effective action is a generic functional of gauge invariant operators
 \be \Gamma= \int d^3x\, \mathcal{L}_{\rm eff}(\mathcal{F}_{\mu \nu}\,,\partial_\mu\mbox{arg}(\Phi_{\rm cl})-a_\mu\,, \psi_{\rm cl} ) + 
 \Gamma_{n.l.}\, ,\label{quantumm-action}\ee
 where we have introduced the field strength $\mathcal{F}_{\mu \nu}=\partial_\mu a_\nu-\partial_\nu a_\mu$ and $\psi_{\rm cl}=|\Phi_{\rm cl}|$.
 The first term in the expression above represents the contribution of {\it local} gauge  invariant terms. $\Gamma_{n.l.}$ depends
 instead on {\it non-local} gauge invariant objects, which may be required if the material has a non-trivial topology.
 Although the action in (\ref{quantumm-action}) is very general, its form simplifies in some limits. This allows us to extract model independent
 properties and, at the same time, identify the predictions of specific models, such as the holographic ones that will be discussed here.

 One of these limits is the small field limit, in which $\Gamma$ is approximated by the Ginzburg-Landau (GL) action:
\be  \Gamma_{\rm \ssGL}= \int d^3x \left(-\frac{1}{4g_0^2}\mathcal{F}_{\mu\nu}^2 -|D_\mu\Phi_{\rm \ssGL}|^2   
+\frac{1}{2\xi_{\rm \ssGL}^2} |\Phi_{\rm \ssGL}|^2-b_{\rm \ssGL}|\Phi_{\rm \ssGL}|^4\right)\, ,\label{GLaction}\ee
where $g_0\, ,\,  \xi_{\rm \ssGL}$ and $b_{\rm \ssGL}$ are real and positive parameters and we have rescaled the scalar field, $\Phi_{\rm \ssGL}\propto \Phi_{\rm cl}$, to have a canonically normalized
kinetic term for $\Phi_{\rm \ssGL}$. Also $D_\mu \Phi_\ssGL=(\partial_\mu-ia_\mu)\Phi_\ssGL$. 
This theory holds if the system is close enough to the symmetry breaking transition, where
 $\Phi_{\rm cl}$ is small.
Another case in which the action simplifies is the limit of slowly varying fields, such that $\Gamma$ can be approximated
by a two-derivative functional 
\be \Gamma\simeq \int d^{3}x\, h(\psi_{\rm
cl})\Big\{-\frac{1}{4g^2(\psi_{\rm
cl})} {\cal F}^2_{\mu \nu}-|D_\mu\Phi_{\rm
cl}|^2   -W(\psi_{\rm
cl})\Big\}\, , \label{limit}
\end{equation}
where $h\, ,\, g$ and  $W$ are unspecified functions of $\psi_{\rm
cl}$. When (\ref{limit}) is a good approximation the SF limit can be taken as $g\rightarrow 0$: this corresponds to taking 
the gauge field non-dynamical and thus the corresponding  U(1) symmetry global. Deep inside a uniform SC, 
where $\psi$ is close to the minimum $\psi_\infty$ of the potential $V=hW$ and the massive gauge field $a_\mu-\partial_\mu$arg$(\Phi_{\rm cl})$
is suppressed, 
(\ref{limit})
reduces to a functional that is  quadratic in the fields; this allows us to define the inverse 
masses of $\psi_{\rm cl}$ and $a_\mu$, respectively $\xi$ and $\lambda$:
\begin{equation}
\frac{1}{\xi^2} = \frac{1}{2 h(\psi_{\infty})} \frac{\partial^2  V}{\partial \psi^2_{\rm cl}}(\psi_\infty) >0\,, \quad \lambda=\frac{1}{\sqrt{2}e(\psi_\infty)\, \psi_\infty}\,.
\label{lambda}
\end{equation}
Note that $\xi$ and $\lambda$ cannot be computed in a model-independent way as their values strongly depend on the 
form of $h\, , \, g$ and $W$. The distinctive property of the superconductive phase is $\psi_\infty \neq 0$, as opposed to the 
normal phase in which $\psi_\infty = 0$. 

Starting from the general action in (\ref{quantumm-action}) it is possible to show in very general terms 
some of the most famous properties of SCs, such as the Meissner effect, the infinite DC conductivity and the 
Josephson effect \cite{Weinberg:1986cq}, which have been  obtained in holography in Refs. 
\cite{Domenech:2010nf,Montull:2012fy,Salvio:2012at,Hartnoll:2008vx,Horowitz:2011dz}.

In this article (unless otherwise stated) we will focus on static configurations, in which everything is constant in time and the temporal 
component of the gauge field is identified with the chemical potential of the U(1) charge: $a_0=\mu$. Also, in order 
to study the thermodynamics of these systems we introduce the free energy
$ F=-T\,  \Gamma$.

In section \ref{Conductivity} we will holographically study the normal phase and explain how to describe  conductors and insulators.
The homogeneous phase, in which $\psi_{\rm cl}=\psi_\infty$, will be studied in section \ref{H-transition} through  the AdS/CFT correspondence. As we 
will see, this provides a rationale for the fact that superconductivity is suppressed at large rather than at small $T$. 

However, inhomogeneous configurations are also of great importance in SCs; e.g. they generically occur when the system
is probed by an external EM field. Among the most famous properties of SCs we can certainly include the existence 
of vortex solutions, which are indeed strongly inhomogeneous configurations. The simplest example of such configurations are straight vortex lines, which can be described by an ansatz
of the form $\Psi_{\rm cl}=\psi_{\rm cl}(r) e^{in\phi}$ and  $a_\phi=a_\phi(r)$ and the other components of the vector potential set to zero. 
Here $r,\phi$ are the usual polar coordinates which parametrize the Euclidean two dimensional space.
By using the action in (\ref{limit}) one obtains the following large $r$ field behavior \cite{Domenech:2010nf}
\begin{equation}
\psi_{\rm cl}\simeq \psi_\infty+\frac{\psi_1}{\sqrt{r}}e^{-r/\xi'}\ , \quad a_\phi\simeq n+a_1\sqrt{r}e^{-r/\lambda'}\, ,
\label{limitrinfty}
\end{equation}
for SCs, while for SFs, where the magnetic field $B=\partial_r a_\phi/r$ is frozen to an external constant  value,
$a_\phi=Br^2/2$, the condensate approaches
the homogeneous value with a power law $\psi_{\rm cl}-\psi_{\infty} \sim n^2/r^2$ \cite{Domenech:2010nf}. 
In Eq. (\ref{limitrinfty}) $\psi_1$ and $a_1$ are unspecified constants, while the action in (\ref{limit}) implies that
$\lambda'=\lambda$ and $\xi'=\xi$. However, the exponential behavior in Eq. (\ref{limitrinfty}) 
tells us  that higher derivative terms cannot be neglected 
as they are generically of the same order of the two-derivative terms. Fortunately, the only effect of the  higher derivatives is to 
modify the values of $\lambda'$ and $\xi'$, such that we generically have $\lambda' \neq \lambda$ and $\xi'\neq \xi$, as discussed in 
\cite{Domenech:2010nf}. $\lambda'$ and $\xi'$ are respectively called penetration depth and coherence length and
are important to characterize SCs. Although the field behavior in (\ref{limitrinfty}) is 
model independent,  $\lambda'$ and $\xi'$, like $\lambda$ and $\xi$,
can only be fixed once the model is specified. Thus their actual values are predictions of the particular model one considers; 
 in section \ref{dynamical-section} we will discuss how to extract them from holography.

While vortex solutions exist for any SC, it is not always the case that there is a range of the external magnetic field $H$ such that
the vortex phase is energetically favorable; when this is true the 
SC is  of Type II and such range is denoted with $H_{c1}\leq H\leq H_{c2}$ (if such range does not exist we have instead a Type I SC). 
$H_{c2}$ is the value of the external magnetic field above which the system is always in the normal
phase.
The vortex 
configuration for Type II SCs and  $H$ slightly smaller than $H_{c2}$ is known to be a triangular lattice of vortices
independently on the specific model one considers: for those high magnetic fields the condensate is small and the GL theory can be  applied 
to predict that configuration \cite{triangular}. $H_{c1}$ is instead the value of $H$ below which the system is in the homogeneous
superconducting phase. It can be computed through the model independent formula\footnote{Here we use 
the normalization of $H$ such that the external current $\vec J_{ext}$ that generates $H$ through $\nabla \times \vec H=g_0^2 \vec J_{ext}$ is coupled
to the  gauge field by means of the interaction term $a_\mu J^\mu_{ext}$ in the Lagrangian.} (see e.g. \cite{Domenech:2010nf})
\be H_{c1}=\frac{g_0^2}{2\pi}(F_1-F_0)\, ,\label{Hc1}\ee
where $F_1$ and $F_0$ are the free energies for the $n=1$ vortex and $n=0$ superconducting phase respectively.
We note that the actual value of $H_{c1}$ is a prediction of the specific model one considers because $g_0$, $F_1$ and $F_0$ are 
model dependent. However, in the  SF limit, $g_0\rightarrow 0$, we always have $H_{c1}\rightarrow 0$; in other words 
$H_{c1}$ is non-trivial only if the magnetic field is dynamical. 

In section \ref{dynamical-section} we will study  the  vortex phase in holography
and illustrate how to compute the critical magnetic fields and show that the holographic SC is of type II. We will emphasize in
particular the differences between the case in which there are no sources of conformal symmetry breaking, other
than $T$ and $\mu$, and the case in which conformal symmetry is broken.

Up to now we have not discussed the role of the non-local terms in Eq. (\ref{quantumm-action}). For reasons which 
will be clear soon, we would like to spend the rest of this introductory section to describe the simplest setup in which $\Gamma_{n.l.}$
is relevant: cylindrical SCs threaded by an external magnetic field along the symmetry axis of the cylinder. To render the discussion
even simpler we will take the deep Type II limit, $g_0\rightarrow 0$, in which the (total) magnetic field coincides with the  external one.
Since there is a non-contractible loop on this geometry we can 
construct non-local gauge invariant objects:
\be W \equiv \exp\left( e i \oint dx^\mu a_\mu\right)\,,\quad m \equiv \frac{1}{2\pi} \oint dx^\mu \partial_\mu \mbox{arg}(\Phi_{\rm cl})= 
\mbox{integer}\, , \quad \dots \  , \ee
where $e$ is the charge of the fundamental charge carriers (say the electrons for real world SCs) and the integrals are performed along the 
non-contractible loop. We will refer to $W$ and $m$ as the Wilson line and the fluxoid respectively and the dots represent other non-local
objects. While the classical action does not depend on $(W, m, \dots)$, quantum effects, such as the Aharonov-Bohm one (or Sagnac one
\cite{anandan,SatoPackard} in the SF literature), could introduce 
a dependence of $\Gamma$ on these quantities. A simple (but still general) way to visualize a dependence of this sort is to think that 
the coefficients of the local part of $\Gamma$ vary with $(W, m, \dots)$: for example, in the domain of validity of the GL theory in 
(\ref{GLaction}), this corresponds to thinking of $\xi_{\ssGL}$ and $b_{\ssGL}$ as functions of $(W, m, \dots)$. 
Such dependence ought to be suppressed in the classical limit and/or when the typical scale of the non-contractible loops, $R$, is large 
compared to the other scales in the problem (such as $1/T$ and $1/\mu$) and therefore we should stay away from these limits to see 
any interest effect. Moreover, {\it since quantum corrections are small in a weakly coupled theory, the biggest effects of the non-local 
quantities are expected in strongly coupled theories}. This justifies the use of holography in this setup (see section \ref{cylinder}). 
Before moving to holography, however, let us identify model-independent effects. 

We shall consider the simplest case in which the magnetic field is constant so that it can be represented by a constant vector potential
along $\chi$,  which parametrizes the compact spatial dimension, $\chi \sim \chi +2\pi R$. Since everything is static and homogeneous an appropriate
ansatz is 
\be \Phi_{\rm cl} = \psi_{\rm cl} e^{i m \chi/R} \, , \quad a=\mu dt+ a_\chi d\chi\, \quad \mbox{with}\,\, \psi_{\rm cl}\,\, \mbox{and}
\,\, a_\chi\,\, \mbox{constant}\, .\label{LPansatz}\ee
Inserting this ansatz in the quantum effective action, Eq. (\ref{quantumm-action}), we obtain
\be \Gamma= \int d^3x\, \mathcal{L}_{\rm eff}(m/R-a_\chi\, , - \mu\,, \psi_{\rm cl} ) + 
 \Gamma_{n.l.}\, \label{quantumm-action-ansatz} ;\ee
in other words, {\it modulo non-local terms}, the system must be periodic in the magnetic flux\footnote{The factor
$1/g_0$ has been introduced in order to have a canonically normalized kinetic term for $a_\mu$ in Eq. (\ref{GLaction}).}
$\Phi(B)\equiv \oint dx^\mu a_\mu/g_0$ with period $\Delta \Phi^{LP}\equiv 2\pi/g_0$, because $\Phi(B)\rightarrow \Phi(B)+\Delta \Phi^{LP}$
can be compensated by a unit shift of the integer $m$. This is known as the Little-Parks effect \cite{LP}. This phenomenon has been observed in experiments, which give  $g_0=2e$, and thus it is
considered as an evidence for Cooper pairing. If the non-local terms in (\ref{quantumm-action-ansatz}) are non-negligible the 
Little-Parks does not generically occur; indeed nothing forces $\Gamma_{n.l}$ to be a function of the local combination $m/R-a_\chi$ only,
but it may depend on $m$ and $W$ separately. The conclusion is that if the system is strongly coupled, far
away from the classical limit and  $R$ is small enough we could see an uplifting of the Little-Parks periodicity to an 
enhanced periodicity, set by the fundamental charge: $2\pi /e$. This would corresponds to resolving 
the internal structure of the {\it composite} condensing operator. Whether  this really occurs is not model-independent.

Remarkably,
we will see (in section \ref{cylinder})  that holography {\it predicts} an uplifting of the Little-Parks periodicity for small	 values of $R$ and that 
this uplifting corresponds in the gravity side to the celebrated Hawking-Page phase transition \cite{Hawking-Page}.

\section{The holographic model}\label{model}
The first step to define a holographic model based on the standard AdS/CFT correspondence is to introduce fields living on an  
asymptotically AdS space. The minimal field content required to describe SCs holographically is a charged scalar field 
$\Psi$ dual to a condensing
operator $\mathcal{O}$ ($\langle \mathcal{O}\rangle=\Phi_{\rm cl}$) and a gauge field $A_\alpha$ dual to the electric current operator $J_\mu$.
As we have stated in the introduction, another ingredient we are interested in is 
a mechanism for conformal symmetry breaking in the IR; this can be achieved by introducing a real scalar $\phi$ (a {\it dilaton}) which acquires 
a non-trivial VEV. The class of actions we consider is \cite{Salvio:2012at}
\be S=\int d^{4}x\, \sqrt{-g}\left\{{{1\over16\pi
G_N}}\left[\mathcal{R}-(\partial_{\alpha} \phi)^2 - V(\phi) \right]\,-\frac{Z_{A}(\phi)}{4g_4^2}{\cal F}_{\alpha \beta}^2-\frac{Z_{\psi}(\phi)}{L^2g_4^2}|D_\alpha\Psi|^2\right\} \, , \label{action} \ee 
where $\mathcal{R}$ is the Ricci scalar, $G_N$ the Newton constant, $V$ the dilaton potential, $g_4$  the U(1) gauge coupling and $L$ is the radius 
of AdS. Also, $Z_{A}$ and $Z_{\psi}$ are generic functions of the dilaton, which do not fulfill particular properties, besides
the fact that they never 
vanish (in order for the semiclassical approximation in the bulk to be justified). Also, $D_\alpha \Psi=(\partial_\alpha -iA_\alpha)\Psi$.

We are interested in describing a static system with two dimensional rotation and translation invariance, for which the metric and the dilaton
have the  form 
\be ds^2=W(z)\left(-f(z)dt^2+dy^2+\frac{dz^2}{f(z)}\right)  , \quad  \phi=\phi(z)\, , \label{generalBH} \ee 
where $dy^2$ is the two dimensional Euclidean metric. The temperature  can be introduced by requiring the presence of a black hole;
so there is a value of the holographic coordinate, $z_0$, such that $f(z_0)=0$. 
Then $T=|f'(z_0)|/4\pi$, where the prime denotes the derivative with respect to $z$. 

Another physical situation 
that can be described with this setup is the case of multiply connected SCs, more precisely of cylindrical geometry: this can be 
simply achieved by compactifying one of the two $y$-coordinates: $\chi \sim \chi +2\pi R$. Then there is necessarily 
another set of solutions \cite{Witten:1998zw}: those obtained from the black holes by exchanging the Euclidean time $t_E=it$ with $\chi$;
these 
configurations have no event horizons and are commonly called AdS solitons. One reason the compactification of $\chi$ is interesting is because this procedure 
allows us to introduce a scale $R$ (other than the $T$ and $\mu$) and thus breaks conformal invariance even in the absence 
of the dilaton. For this reason from now on the case in which $\chi$ is compact will be discussed with a simplified field content, where 
$\phi$ is removed and $Z_A=Z_\psi=1$. The AdS soliton (black hole) geometry is energetically favorable with respect to the black hole (AdS soliton)
at sufficiently small (large)
 temperatures, $T\leq 1/(2\pi R)$ ($T\geq 1/(2\pi R)$). Another reason why we are interested in compactifying a spatial coordinate is to study what happens when cylindrical 
SCs are threaded by magnetic fields: in these physical setups indeed a universal prediction of holography emerges, as 
it will be discussed in section \ref{cylinder} and anticipated in the introduction.

Although we include the possibility of conformal symmetry breaking in the IR, in this article we will always assume that the ultraviolet (UV) is 
conformally invariant, or in other words that the fields are approximated by an AdS configuration close to a value of $z$, say $z=0$: 
for $z$ close to zero,  $W(z)\simeq L^2/z^2$, $f(z)\simeq 1$ and $\phi(z)\simeq 0$. Thus we can use the standard AdS/CFT dictionary, 
which relates the properties of a gravitational theory with those of a CFT. In particular the $z=0$ 
values of $\Psi$ and $A_{\mu}$ are the sources of $\mathcal{O}$ and $J_{\mu}$ in the CFT. 
Therefore, if one solves the bulk field equations with boundary conditions 
\be \Psi|_{z=0}=\Psi_0\,, \quad A_{\mu}|_{z=0}=a_{\mu} \, , \label{s-amu}\ee
 the Green's function for $\mathcal{O}$ and $J_{\mu}$ are given by differentiating
 the on-shell action with respect to $\Psi_0$ and $a_{\mu}$; e.g. the vacuum expectation values are given by
\be 
\langle  J_\mu\rangle=\frac{1}{g_4^2}{\cal F}_{z\mu}|_{z=0}\ , \quad  \langle{\cal
O}\rangle=\frac{1}{g_4^2 z^{2}}D_z \Psi|_{z=0}\, .
\label{operator} \ee

In order to keep the discussion as simple as possible we assume that $\Psi$ and $A_\alpha$ do not backreact on the geometry and the 
dilaton; this can be consistently achieved by taking the limit $G_N\rightarrow 0$. A generalization to the case $G_N\neq 0$ can be found in 
\cite{Hartnoll:2008kx,Horowitz-Way}. 
A general form of asymptotically AdS dilatonic black hole solutions
with the assumed symmetries was derived in \cite{Anabalon:2012ta}. One important aspect of these black holes is that the  dilaton
is not generically constant but runs with the  energy:
\be \phi(z)=  \sqrt{\frac{\nu^2-1}{2}}\ln(1+z/L)\, , \label{running} \ee
where $\nu$ is a real parameter of the potential such that $\nu \geq 1$ and  $\nu=1$  recovers the Schwarzschild black hole (S-BH) of Einstein's gravity.
Eq. (\ref{running}) tells us that although the theory has a UV fixed point, $\phi(0)=0$, conformal invariance is broken and maximally in the IR, $z=z_0$.
In order for the configurations of Ref. \cite{Anabalon:2012ta} to be a solution, $V$ has to be appropriately chosen. In particular
the requirement of an asymptotically AdS configuration space with cosmological constant $\Lambda$ implies that at the conformal point
(conventionally $\phi=0$) where $V(0)=\Lambda$ we have $\partial V/\partial \phi=0$. 
As a result, for each value of $\nu$, even $\nu \neq 1$,   the S-BH with cosmological constant $\Lambda$ is always a solution.
One finds \cite{Salvio:2012at} that the S-BH is favorable at high temperatures, while the dilaton-BH, Eq. (\ref{running}), dominates in the 
low temperature region.

\section{Conductivity}\label{Conductivity}
Having described the solutions of the dilatonic gravity systems we now want to understand the type of materials they correspond to.
Some information can be gained by studying the conductivity, $\sigma$. This will also shed light on the nature of the SF  phase
transitions, which will be discussed in the next section.

To compute $\sigma$ let us consider, on top of these geometries, a small plane wave along a spatial coordinate $x$, 
\be A_x (t,z)=\mathcal{A}(z)e^{i\omega (p(z)-t)}\,, \label{wave-decomposition}\ee
 which is induced by a small electromagnetic field, $A_x(t,z)|_{z=0}=a_x (t)$. Here $\mathcal{A}$ and $p$ are real functions of $z$. 
 The system responds creating a current which is linear in the electric field $E_x$: that is  $\langle J_x\rangle= \sigma E_x$.  
Using the AdS/CFT dictionary, the first equation in (\ref{operator}), we have
\be \sigma=  \frac{p'(0)}{g_4^2}-i\frac{ \mathcal{A}'(0)}{g_4^2\omega \mathcal{A}(0)}\, . \label{sigma}\ee

When an event horizon is present regularity of the solution implies that the plane wave should be ingoing rather than outgoing from the 
horizon. It can be shown that this results in a non-vanishing Re$[\sigma]$ and DC conductivity \cite{Salvio:2012at}
\be \lim _{\omega \rightarrow 0} \mbox{Re}[\sigma]=\frac{1}{g_4^2}Z_A|_{z=z_0} \neq 0 \, . \label{DCsigma}\ee
For an AdS soliton, which has  no event horizon, regularity allows for a vanishing DC conductivity, which has been identified with 
an insulating behavior \cite{Nishioka:2009zj,Horowitz-Way}. 
Therefore, compactifying a spatial dimension $\chi$ allows us to realize a conductor/insulator
transition as the temperature is lowered. Eq. (\ref{DCsigma}), however, tells us that there is another way to suppress the DC conductivity
and obtain such transition if a dilaton is present
in the  spectrum: if one chooses $Z_A$ sufficiently small for large values of $\phi$, the running in Eq. (\ref{running}) implies that
$\lim _{\omega \rightarrow 0}$Re$[\sigma]$ is small, especially in the low temperature limit where $z_0$ is large. 
In this setup the transition (discussed in the previous section) between the S-BH and dilaton-BH,
which is obtained by lowering the temperature, represents a conductor/insulator transition. It is worth noting 
that insulating systems corresponds to solids in the  fluid mechanical interpretation \cite{Montull:2012fy}.

\section{Superfluid phase transition}\label{H-transition}
We now move on and study the simplest example of superconducting state: a static, homogeneous and isotropic material
in which the U(1) symmetry is spontaneously broken. The ansatz is 
\be \Psi=\psi(z)\,,\quad A=A_0(z)dt \, .\label{ansatz}\ee
The reader may wonder why we have introduced the temporal component of the gauge field, because having a non-vanishing
$\psi(z)$ seems already enough to describe the state we are interested in. The reason is that no regular solution is found 
when $A_0=0$  in (\ref{ansatz}), as shown in \cite{Salvio:2012at}, and we want to exclude singular solutions. 
Physically this corresponds to the fact that there is only 
one parameter breaking 
conformal symmetry ($T$ for the black hole and $R$ for the AdS soliton) in the absence of the dilaton\footnote{The dilaton
cannot change this conclusion as it is regular for any $T$ and the function $Z_\psi$, Eq. (\ref{action}), is assumed to be regular
and non-zero.}, thus no phase transition can occur 
in this case. 
Therefore we set the UV boundary conditions
\be \Psi_0=0\,, \quad a_0=\mu \,. \label{UV}\ee
The first condition guarantees that the U(1) symmetry is spontaneously broken, while the second one, with a non-vanishing chemical potential
$\mu$, keeps the profile of $A_0$ different from zero and allows us to find a regular solution.

An immediate consequence of this argument is that 
when the  temperature is large compared to the chemical potential a black hole suppresses the superconductivity. At small temperatures
instead the system turns into a SC \cite{Hartnoll:2008vx}. In AdS/CFT there is therefore a reason for the fact that a metal becomes a SC
at small rather than at large temperatures. For the AdS soliton the same conclusion can be reached, but with $T$ substituted by $1/R$.

According to the results of the previous section these SF phase transitions can be either of the conductor/SC or 
insulator/SC type, depending on the behavior of the conductivity in the normal phase:  the 
two cases correspond to  a 
non-vanishing or vanishing DC conductivity respectively. In the fluid mechanical interpretation, the SF phase
of a holographic system with a solid normal phase has been identified with a supersolid \cite{Montull:2009fe}.

\section{Dynamical gauge fields in AdS/CFT and superconductivity} \label{dynamical-section}

In the holographic results we have discussed so far the dynamics of the EM field is not important. Therefore, according to the 
discussion presented in the introduction, they can be applied equally well to SFs and SCs. In this section
instead we discuss some important effects of superconductivity, which crucially rely on the dynamics of the EM field.

However, we  note that imposing the first boundary condition in (\ref{s-amu}) treats $a_\mu$ as an external source,
which does not participate in the dynamics
of the system. In order to have a dynamical $a_\mu$ we should integrate over all possible field configurations:  (working in the 
Euclidean space) 
\be
Z[J_{ext}]=\int Da\,  e^{-S_\ssE[a_\mu]+\int d^{3}x \left(-\frac{1}{4g_b^2}{\cal F}^2_{\mu \nu}+a_\mu J^\mu_{ext}\right)}\, ,
\label{newg}
\ee
where $S_\ssE[a_\mu]$ is the Euclidean bulk action computed on a solution of the field equations with  boundary condition
in  (\ref{s-amu}). Also, we have introduced for generality a kinetic term for $a_\mu$ and an external current 
$J_{ext}$. Then $Z[J_{ext}]$ defines as usual the generating functional for the Green functions of $a_\mu$.
Of course, if there are other operators in the theory, besides the gauge field,  $S_\ssE[a_\mu]$ in Eq. (\ref{newg}) 
will depend on the corresponding sources as well; for example in the case of superconductivity discussed in the previous sections
$S_\ssE[a_\mu]$ also depends on $\Psi_0$, which we introduced in (\ref{s-amu}).

In the semiclassical limit this procedure reduces to solving the Maxwell equations,
\be
\langle  J^\mu\rangle+\frac{1}{g_b^2}\partial_\nu {\cal F}^{\nu \mu}+J^\mu_{ext}=0\, ,
\label{maxwell}
\ee
where we have used  $\langle \hat J^\mu\rangle=-\delta S_\ssE/\delta a_\mu$. 
The semiclassical limit corresponds to taking a large external current, $g_b\rightarrow 0$ and  a limit 
on the parameters of the bulk theory such that $S_\ssE[a_\mu]$ becomes large; for example, for the theory defined in (\ref{action})
this limit is $G_N\rightarrow 0$ and $g_4\rightarrow 0$.

Now, using  Eq.~(\ref{operator}), we can see that the Maxwell equations in (\ref{maxwell}) can viewed as a Neumann type
boundary condition in the gravity side:
\be
\frac{1}{g_4^2}{\cal F}_z^{\,\,\, \mu} \Big|_{z=0} +\frac{1}{g_b^2}\partial_\nu {\cal F}^{\nu \mu}\Big|_{z=0}+J^\mu_{ext}=0\, .
\label{maxwell2}
\ee
The lesson is therefore that switching from the Dirichlet boundary condition (\ref{s-amu}) to the condition above promotes 
$a_\mu$ to a dynamical field.

As far as the holographic SC model of section \ref{model} is concerned, this procedure has been applied to find the Meissner effect 
and genuine SC vortices \cite{Domenech:2010nf,Salvio:2012at}, at least in the simplest case of a straight vortex line:
\be \Psi=\psi(z,r)e^{i n\phi}\  ,\quad A_0=A_0(z,r)\ , \quad
A_{\phi}=A_{\phi}(z,r)\, . \label{ansatz-vortex} \ee
$r,\phi$ are the usual polar coordinate parametrizing the two dimensional Euclidean space and so here 
we assume that there are at least two non-compact dimensions in the CFT. This is not the case for the four dimensional AdS soliton;
however, vortex solutions have been found on top of the five dimensional AdS soliton in \cite{Montull:2012fy}.
Using the ansatz in (\ref{ansatz-vortex}) the Neumann-like boundary condition in (\ref{maxwell2}) becomes 
\be
\frac{1}{g_4^2}\partial_z A_\phi\Big|_{z=0} +\frac{1}{g_b^2}r\partial_r\left(\frac{1}{r}\partial_r  A_\phi\right)\Big|_{z=0}+J_{ext\, \phi}=0\label{newbcsc}\, ,
\ee
while the requirement of spontaneous symmetry breaking and the presence of a finite charge
density again fixes the other $z=0$ boundary conditions, Eq. 
(\ref{UV}). Regularity of the solutions instead fixes the conditions at $z=z_0$ and at the center of the vortex $r=0$ and physical 
conditions on the behavior at infinity, $r\rightarrow \infty$, imposes constraints on the remaining boundary.

This results in genuine SC profiles for the total magnetic field $B=\partial_r a_\phi/r$: for example for $n=0$ one recovers the Meissner
effect, while for $n\neq 0$ one observes the exponential damping of $B$ far away from the center of the vortex 
\cite{Domenech:2010nf,Montull:2012fy,Salvio:2012at}, Eq. (\ref{limitrinfty}), allowing for a holographic prediction for $\lambda'$ and 
$\xi'$. Such properties should be contrasted
with the Dirichlet boundary condition for the magnetic field, which in polar coordinates is $A_{\phi}|_{z=0}=Br^2/2$ (with $B$ constant), corresponding
to SF systems (see e.g. \cite{Montull:2009fe,Domenech:2010nf,Montull:2012fy,Salvio:2012at,Roychowdhury:2012hp}).

Once the vortex solutions are obtained we can compute $H_{c1}$ through Eq. (\ref{Hc1}). Indeed selecting  the $n=1$ vortex and $n=0$ homogeneous
superconducting phase we can compute $F_1-F_0$ while, integrating over the bulk {\it \`a la} Kaluza-Klein gives $g_0$:
\be \frac{1}{g_0^2}= \frac{1}{g_4^2}\int_0^{z_0}dz\, Z_A(\phi)\, . \label{KK} \ee
The fact that the dilaton appears in this formula has an important consequence on the behavior of $H_{c1}$ at low temperatures
\cite{Salvio:2012at}, as we now explain. In the 
absence of $\phi$  Eq. (\ref{KK}) implies that $g_0\rightarrow 0$ as $T\rightarrow 0$ (in this limit the horizon is removed, $z_0\rightarrow 
\infty$) and Eq. (\ref{Hc1}) forces $H_{c1}$ to vanish. The physical reason why this emerges is because 
the theory without the  dilaton is scale invariant and so $g_0$ should go to zero as $T\rightarrow 0$. The presence of the dilaton breaks 
scale invariance and can avoid this conclusion, providing a non-vanishing $H_{c1}$: $g_0$ remains non-zero all the way down to $T=0$ if $Z_A(\phi)$ goes to zero 
fast enough as $\phi\rightarrow \infty$, a limit that always occurs close to the horizon as $T\rightarrow 0$, Eq. (\ref{running}).
Interestingly, when $H_{c1}$ remains non-zero at zero temperature, Eq. (\ref{sigma}) tells us that the  corresponding normal phase  must have 
a suppressed DC conductivity in the low temperature region, resembling an insulator.

The second critical field $H_{c2}$ can be instead computed as the value of $H$ at which the condensate goes to zero. Up to now
all the holographic models of superconductivity have turned out to be of Type II \cite{Domenech:2010nf,Montull:2012fy,Salvio:2012at},
like all known high temperature SCs.

 \section{Multiply connected holographic superconductors and superfluids} \label{cylinder}
As explained in section \ref{introduction}  an interesting setup to apply the holographic techniques are multiply connected SCs probed 
by magnetic fields. We shall consider the simplest case described at the end of section \ref{introduction}. In the presence 
of a compactified dimension scale invariance is broken and we will assume for simplicity that the bulk field content does
not include the dilaton. Also, the ansatz in (\ref{LPansatz}) corresponds holographically to 
\be \Psi= \psi(z)e^{im\chi/R}\, , \quad A_0=A_0(z)\, , \quad A_\chi=A_\chi(z) \, , \ee
with UV boundary conditions $\Psi|_{z=0}=0\, ,\, A|_{z=0}=\mu\, , \, A_{\chi}|_{z=0}=a_\chi$. As already discussed, the system is in the S-BH phase at large radii and in the AdS soliton otherwise and, as shown
in \cite{Montull:2011im}, such transition corresponds to an uplifting of the  Little-Parks periodicity. This is a universal prediction
of holography as it  uniquely relies on the  geometrical properties of the  S-BH and the  AdS soliton. 
The simplest way to understand this point is to look at the IR boundary condition (at $z=z_0$) for $A_\chi$ and $\psi$,
which regularity of the  bulk solutions requires:
\be\left.A_\chi' + {2\,\psi^2\over 3\, z_h} \left(A_\chi-{ m\over R}\right)\right|_{z=z_h}=
\left.\frac{3\psi'}{z_0}  + \left( A_\chi-\frac{m}{R}\right)^2 \psi \right|_{z=z_0}
  =0\label{BHBCs}\ee
for the S-BH and 
\be A_\chi|_{z=z_0}=0\, , \quad \bac \left\{\bac
\left.\frac{3\psi'}{z_0}- A_0^2 \psi\right|_{z=z_0}=0 \quad {\rm for~}m = 0 \ , \\ \phantom{v} \\ 
\qquad\quad\, \psi |_{z=z_0} = 0 \quad{\rm for~} m \neq 0\, \ea\right.\ea \label{solitonBCs}\ee
for the AdS soliton. While the conditions in (\ref{BHBCs}) only depend on the local combination of $m$ and $W$, that is $m/R-a_\chi$, the conditions
in (\ref{solitonBCs}) depend separately on the fluxoid and the Wilson line. This means that the non-local terms in the quantum effective action
are suppressed (unsuppressed) in the  S-BH (AdS soliton) phase. Correspondingly, according to the  model-independent discussion
of section \ref{introduction}, the system should be characterized by the Little-Parks periodicity for the S BH and an uplifting of such periodicity
for the AdS soliton as shown in \cite{Montull:2011im,Montull:2012fy}.

\section*{Acknowledgements} We would like to thank Oriol Dom\`enech, Marc Montull, Alex Pomarol, Oriol Pujol\`as and Pedro J. Silva for
collaborations and Daniele Dorigoni and Massimo Mannarelli for useful discussions. This work was partly 
supported by the EU ITN ``Unification in the LHC Era", contract PITN-GA-2009-237920 (UNILHC) 
and by MIUR under contract 2006022501.


\section*{References}
\begin{thebibliography}{99}

\bibitem{Hartnoll:2009sz}
  Hartnoll S A 2009
  {\it Class. Quant. Grav.}  {\bf 26}  224002
  ({\it Preprint} arXiv:0903.3246 [hep-th]).
  Herzog C P 2009
  {\it J. Phys. A} {\bf 42} 343001
  ({\it Preprint}  arXiv:0904.1975 [hep-th]).
  McGreevy J 2010
  {\it Adv. High Energy Phys}  723105
  ({\it Preprint} arXiv:0909.0518 [hep-th]).
 Hartnoll S A 2009
  {\it Preprint} arXiv:0909.3553 [cond-mat.str-el].
  Sachdev S 2010
  {\it Preprint} arXiv:1002.2947 [hep-th].
  Pires A S T 2010
  {\it Preprint} arXiv:1006.5838 [cond-mat.str-el].
  Horowitz G T 2011
  {\it Class. Quant. Grav.}  {\bf 28}  114008
  ({\it Preprint} arXiv:1010.2784 [gr-qc]).
  Hartnoll S A 2011
  {\it Preprint} arXiv:1106.4324 [hep-th].
  
  \bibitem{doping} Lee P A,  Nagaosa N and Wen X 2006 
  {\it Rev. Mod. Phys.} 78 17$-$85.
  
\bibitem{Hartnoll:2008vx}
  Hartnoll S A, Herzog C P and Horowitz G T 2008 {\it Phys. Rev. Lett.} {\bf 101}  031601
  ({\it Preprint} arXiv:0803.3295 [hep-th]).
  %
  
  
\bibitem{Domenech:2010nf}
  Dom\`enech O, Montull M, Pomarol A, Salvio A and Silva P J 2010
  {\it JHEP} {\bf 1008}  033
  ({\it Preprint} arXiv:1005.1776 [hep-th]).
 
\bibitem{Herzog:2008he}
  Herzog C P, Kovtun P K and Son D T 2009
  {\it Phys. Rev. D} {\bf 79}  066002
  ({\it Preprint} arXiv:0809.4870 [hep-th]).


\bibitem{Hartnoll:2009ns}
  Liu H, McGreevy J and Vegh D 2011
  {\it Phys. Rev. D} {\bf 83}  065029
  ({\it Preprint}  arXiv:0903.2477 [hep-th]).
  Hartnoll S A, Polchinski J, Silverstein E and Tong D 2010
  {\it JHEP} {\bf 1004} 120
  ({\it Preprint} [arXiv:0912.1061 [hep-th]).
  Pal S S 2012
  {\it Preprint} arXiv:1209.3559 [hep-th].
  
\bibitem{Weinberg:1986cq}
  Weinberg S 1986
  {\it Prog. Theor. Phys. Suppl.}  {\bf 86}  43.
  
  
\bibitem{Montull:2012fy}
  Montull M, Pujol\`as O, Salvio A and Silva P J 2012
   {\it JHEP} {\bf 1204}  135
  ({\it Preprint} arXiv:1202.0006 [hep-th]).

  
\bibitem{Salvio:2012at}
  Salvio A  2012 {\it JHEP} {\bf 1209} 134 ({\it Preprint} arXiv:1207.3800 [hep-th]).
  
  
  
  
\bibitem{Horowitz:2011dz}
 Horowitz G T, Santos J E and Way B 2011
  {\it Phys. Rev. Lett.}  {\bf 106}  221601
  ({\it Preprint} arXiv:1101.3326 [hep-th]).

  
\bibitem{triangular}
Kleiner W H, Roth L M and Autler S H 1964 {\it Phys. Rev. A} {\bf 133}
 1226.
 
 
\bibitem{anandan}
Anandan J 1981 {\it Phys. Rev. Lett.} 47, 7.

\bibitem{SatoPackard}
Sato Y and Packard R E 2012 {\it Rep. Prog. Phys.} 75 016401.  



\bibitem{LP}
Little W A, Parks R D 1962 {\it Phys. Rev. Lett.} {\bf 9}, 9.


 \bibitem{Hawking-Page}
 Hawking S W, Page D N 1983 {\it Commun. Math. Phys.} {\bf 87}, 577.
 
  

\bibitem{Witten:1998zw}
  Witten E 1998 {\it Adv. Theor. Math. Phys.} {\bf 2 } 505-532
  ({\it Preprint} hep-th/9803131). Horowitz G T and Myers R C 1998
   {\it Phys. Rev.}  {\bf D59 }   026005 ({\it Preprint} hep-th/9808079).
  
  
\bibitem{Hartnoll:2008kx}
  Hartnoll S A, Herzog C P and Horowitz G T 2008
  {\it JHEP} {\bf 0812} 015
  ({\it Preprint} arXiv:0810.1563 [hep-th]).
     
   

   
\bibitem{Horowitz-Way}
  Horowitz G T and Way B 2010 {\it JHEP} {\bf 1011}  011
  ({\it Preprint} arXiv:1007.3714 [hep-th]).
  

  

  
\bibitem{Anabalon:2012ta}
  Anabalon A 2012 {\it JHEP} {\bf 1206} 127 ({\it Preprint} arXiv:1204.2720 [hep-th]).
  
\bibitem{Nishioka:2009zj}
  Nishioka T, Ryu S and Takayanagi T 2010
  {\it JHEP} {\bf 1003}  131 ({\it Preprint} arXiv:0911.0962 [hep-th]).
  
  
  
  
\bibitem{Montull:2009fe}
  Montull M, Pomarol A and Silva P J 2009
  {\it Phys. Rev. Lett.}  {\bf 103}  091601
  ({\it Preprint} arXiv:0906.2396 [hep-th]).
    Keranen V, Keski-Vakkuri E, Nowling S and Yogendran K P 2010
  {\it Phys. Rev. D} {\bf 81} 126012
  ({\it Preprint} arXiv:0912.4280 [hep-th]).

\bibitem{Roychowdhury:2012hp}
  Roychowdhury D 2012
  {\it Phys. Rev. D} {\bf 86}  106009
  ({\it Preprint} arXiv:1211.0904 [hep-th]).
  Roychowdhury D 2013
  {\it Phys. Lett. B} {\bf 718}  1089
  ({\it Preprint} arXiv:1211.1612 [hep-th]).
  
\bibitem{Montull:2011im}
  Montull M, Pujol\`as O, Salvio A and Silva P J 2011
  {\it Phys. Rev. Lett.}  {\bf 107}  181601
  ({\it Preprint} arXiv:1105.5392 [hep-th]).

  
\end{thebibliography}
\end{document}